\documentclass[conference]{IEEEtran}
\ifCLASSINFOpdf
  \usepackage[pdftex]{graphicx}
  \DeclareGraphicsExtensions{.pdf,.jpeg,.png,.eps}
\else
  \usepackage[dvips]{graphicx}
  \DeclareGraphicsExtensions{.eps}
\fi
\usepackage[cmex10]{amsmath}
\usepackage{amssymb,amsthm}
\usepackage{mathtools, cuted}
\usepackage{lipsum, color}
\usepackage[utf8]{inputenc}
\usepackage[left=0.5in,right=0.5in,bottom=0.65in,top=0.65in]{geometry}
\usepackage[tight,footnotesize]{subfigure}
\newcommand{\argmin}{\operatornamewithlimits{arg\,min}}

\begin{document}
%
\title{Message Authentication and Secret Key Agreement in VANETs via Angle of Arrival}
%
%
%
%
\author{\IEEEauthorblockN{Amr Abdelaziz$^*$, Ron Burton$^\dagger$, and C. Emre Koksal$^*$} \IEEEauthorblockA{$^*$Department of Electrical and Computer Engineering, The Ohio State University\\
$\{$abdelaziz.7, koksal.2$\}$@osu.edu\\
$^\dagger$Transportation Research Center Inc, Columbus, OH\\
Burtonr@trcpg.com\\
}}
\maketitle

\begin{abstract}
In the scope of VANETs, nature of exchanged safety/warning messages renders itself highly location dependent as it is usually for incident reporting. Thus, vehicles are required to periodically exchange beacon messages that include speed, time and GPS location information. In this paper paper, we present a physical layer assisted message authentication scheme that uses Angle of Arrival (AoA) estimation to verify the message originator location based on the claimed location information.
 \par
Within the considered vehicular communication settings, fundamental limits of AoA estimation are developed in terms of its Cramer Rao Bound (CRB) and existence of efficient estimator. The problem of deciding whether the received signal is originated from the claimed GPS location is formulated as a two sided hypotheses testing problem whose solution is given by \textit{Wald test statics}. Moreover, we use correct decision, $P_D$, and false alarm, $P_F$, probabilities as a quantitative performance measure. The observation posterior likelihood function is shown to satisfy regularity conditions necessary for asymptotic normality of the ML-AoA estimator. Thus, we give $P_D$ and $P_F$ in a closed form.
\par 
 We extend the potential of physical layer contribution in security to provide physical layer assisted secret key agreement (SKA) protocol. A public key (PK) based SKA in which communicating vehicles are required to validate their respective physical location. We show that the risk of the Man in the Middle attack, which is common in PK-SKA protocols without a trusted third party, is waived up to the literal meaning of the word "middle".

\end{abstract}


%
\IEEEpeerreviewmaketitle

\section{Introduction}
Security of wireless VANET is of great importance due to the close relation between the information exchanged in VANETs to the public safety. The untethered nature of the open wireless medium of VANETs open the door for a wide range of security vulnerability issues. Message authentication, maintaining privacy, confidentiality, non-repudiation and information integrity are all basic security requirements in a typical communication network. In addition, the high mobile nature of VANET users together with the sensitive safety information exchanged require real time availability of the network. Therefore, the stringent delay requirements in VANETS impose further constraints on the complexity of the potential solutions. These requirements, in fact, offers a big challenge due to the large scale of VANET. Moreover, the cooperative nature of VANETs offers a set of an additional security challenges. That is the impact of security openings can propagate to the entire network due to frequent exchange of messages among all vehicles. Therefore, message source authentication is a major crucial requirements in VANETs as bogus messages can threaten drivers safety and/or convenience. For example in a highway, false warning or safety message injected from outsider or an insider with rational intentions may cause the entire traffic to be blocked and may have potentially fatal safety threatening consequences.

\par
There is an extensive research made to assess security vulnerabilities and requirements of  VANET. Denial of service attacks including flooding, jamming \cite{jamming_vul}, spamming, malware and black hole attacks have been reported to have a potential effect up to complete blockage of VANET. Moreover, attacks on node authentication is another type of attacks that is of a great concern as an unauthorized activity may threaten people lives. Message authentication in VANETs is based on public key infrastructure (PKI) where a certificate authority (CA) is responsible for issuing, manage, distribute and revoke certificates \cite{VANET_SEC_BOOK}. The use of authentication certificate may eliminate the possibility of impersonation attack, however, it offers a considerable message overhead as it requires the certificate to be associated with each message transmission \cite{overhead}. Moreover, it doesn't eliminate the possibility of using a stolen identity (Masquerading Attack), message retransmission (Reply Attack \cite{Replay}-can be mitigated by associating an authenticated time stamp at each message transmission) or the use of multiple legitimate identities (Sybil Attack \cite{sybil}) by an illegitimate node. Several solutions have been proposed to mitigate the potential of the aforementioned attacks. Readers may refer to \cite{survey} for an extensive survey on security in VANETs.
\par
Security at the physical layer is the information theoretic counterpart of the computational security designed in the upper communication layers. At the physical layer, several parameters like Received Signal Strength (RSS) \cite{RSSI}, Channel State Information (CSI) \cite{key_csi}, Angle of Arrival (AoA) \cite{securearray} and Angle of Departure (AoD) can be exploited to provide reliable and time efficient security tools. The nature of messages exchanged over VANETs, usually reporting incident at a given location, renders itself highly location dependent. Thus, VANET nodes are required to exchange beacon messages periodically that incorporates user specific information including anonymous identity, time, GPS location and speed to ensure the cooperative awareness of neighboring vehicles. An illegitimate node may intentionally falsify these information as to achieve a certain goal which might be rational in some scenarios. Therefore, in contrast to other physical layer parameters, AoA is of a contextual meaning that can contribute to message authentication decision. 


\par
 In this paper paper, we present a physical layer assisted message authentication scheme that uses AoA estimation to verify the message originator location. The scheme makes use of the information contained in beacon messages to validate the claimed GPS location information with the AoA information obtained at the physical layer. The proposed scheme offers a physical cross verification tool that integrates the existing conventional PKI message authentication and makes use of the available physical layer information. The proposed scheme comes as a solution for message authentication problem in VANETs under the following impersonation/message substitution attack scenarios:
\begin{enumerate}
\item Attacker with single or multiple stolen identities (Sybil attack) that intentionally falsies different GPS locations.
\item Attacker that spoofs/jams GPS information \cite{gps_spoof},\cite{gps_spoof_2} of other nodes in the network to let them report their incidents in certain/erroneous GPS locations. 
\end{enumerate}
\par
Noting that in all of the considered attack scenarios, message received at the target receiver will pass the conventional PKI procedures as it is either originated from an attacker with a stolen identity or a legitimate transmitter with spoofed/jammed GPS information. To the best of the authors knowledge, no solutions are available at the upper security layer to such types of attacks.

\par
To that end, we develop the fundamental limits of AoA estimation in terms of its Cramer Rao Bound (CRB) which sets a lower bound on any AoA estimator variance. We also proceed by introducing the maximum likelihood (ML) AoA estimator which is shown to be consistent asymptotically in the array size. Based on the knowledge of the declared AoA of arrival of all neighboring vehicles/Road side units (RSU) extracted from the beacon messages, $\theta_b$, we formulate the problem of deciding whether the received signal is originated from the declared physical direction of a given vehicle/RSU  as a two sided hypotheses testing problem whose solution is given by Wald test statics \cite{wald}. Moreover, as a quantitative measure for the proposed scheme performance, we use probability of denying an authentic transmission ($1-P_D$ with $P_D$ being the probability of correct decision) and the probability of accepting an illegitimate message (false alarm probability $P_F$). We show that the observation posterior likelihood function satisfies regularity conditions necessary for the asymptotic normality of the ML-AoA estimator. Thus, we find $P_D$ and $P_F$ in a closed form.
\par
In addition, another major challenge that faces the prevalence of VANETs is the secret key agreement(SKA) between peer vehicles. In the scope of VANETs, secure interaction between vehicles should be established rapidly without any interaction with the CA during the session key exchange. The process of sharing the secret key is called Secret Key Agreement (SKA). In this paper, we extend the potential of physical layer contribution in security to provide a novel physical layer assisted secret key agreement protocol. A public key based SKA in which communicating vehicles are able to validate their respective physical location based on the claimed location information in the beacon messages. We show in an algorithmic way that risk of the Man in the Middle (MitM) attack, which is common in PKI SKA protocols with no trusted third party, is waived up to the literal meaning of the word "middle". 
\par
Contributions of this work can summarized as follows:
\begin{itemize}
\item We introduce the AoA information as a security parameter that is used in conjunction with conventional PKI as solution to the problem of security against attacker with stolen identities.
\item We develop the fundamental limits of AoA estimation for the considered modelin terms of CRB, also, we provide a closed form expression for the correct decision/false alarm probabilities as a quantitative measure for the proposed scheme performance.
\item We introduce a novel joint SKA protocol that turns the MitM attack into the literal meaning of the word "middle" (the attacker has to be located in the middle).  
\end{itemize}
\par
\textbf{Related Work.} This not the first work that address physical layer exploitation for enhanced security in VANET. In \cite{phy_sec_vanet}, a Physical layer Assisted message Authentication (PAA) under PKI in vehicular communication networks in which a trust between two vehicles can be maintained by comparing the current estimated channel response and the previous estimated channel response. This method offers a tool for maintaining authentication between two communication nodes, however, the initial authentication phase is still vulnerable. Rather, the estimated channel response by itself is contextually meaningless, \textbf{i.e.}, it can not provide any information that can be used by the application layer for further assessment of the message contents.

\section{System and Attack Models}
\label{sec:model}
In the rest of this paper we use boldface uppercase letters for random vectors/matrices, uppercase letters for their realizations, bold face lowercase letters for deterministic vectors and lowercase letters for its elements. While, $(.)^{*}$ denotes conjugate of complex number, $(.)^{\dagger}$ denotes conjugate transpose, $\mathbf{I}_N$ denotes identity matrix of size $N$, $\mathbf{tr}(.)$ denotes matrix trace operator, $\textbf{var}(.)$ denotes variance of random variable, $\mathbb{E}[.]$ dentoes expectation operator, $\det(.)$ denotes matrix determinant operator and $\mathbf{1}_{m \times n}$ denotes a $m \times n$ matrix of all 1's. 
\subsection{System model}
\label{subsec:sysmodel}
As illustrated in Fig. (\ref{fig:sysmodel}), we consider the scenario in which a VANET consists of multiple vehicles in the vicinity of a RSU of known location. Beacon messages are exchanged periodically to collect neighboring vehicles GPS locations, time and speed information. An attacker with a stolen identity falsifies its location information aiming to mislead the target vehicles to accept its message. The stolen identity may be for another vehicle or RSU. For simplicity, we assume all communication nodes to be equipped with multiple antenna transceivers each of array size $n$, however, the case of different array size at each node does not affect the obtained results. Similarly, we will assume the uniform linear array (ULA) antenna configuration, however, the obtained results apply directly to any other antenna configuration with straightforward manipulation.
\par 
The attacker message, $\mathfrak{M}$, is divided into $n_p$ packets, each packet is sent over the air using 802.11p \cite{802_11_STD} physical layer in the form of $n_s$ OFDM symbols. Each OFDM symbol consists of 64-subcarriers, among the 64 subcarriers, 52 are used for data transmission, which is further composed of 48 data and 4 pilot subcarriers. The pilot subcarriers are usually used for channel estimation, however, in our settings we use it for AoA estimation purpose as well. Thus, no extra communication overhead is needed for the AoA estimation. Further, over each subcarrier, we assume the channel to be flat, i.e., the the coherence bandwidth is larger than the bandwidth of each subchannel. The discrete baseband equivalent channel (after FFT operation at the receiver) for the signal received by one of the target vehicles can be expressed as:
\begin{figure}[!tb]
\centering
\includegraphics[width=3.4 in, height = 2 in]{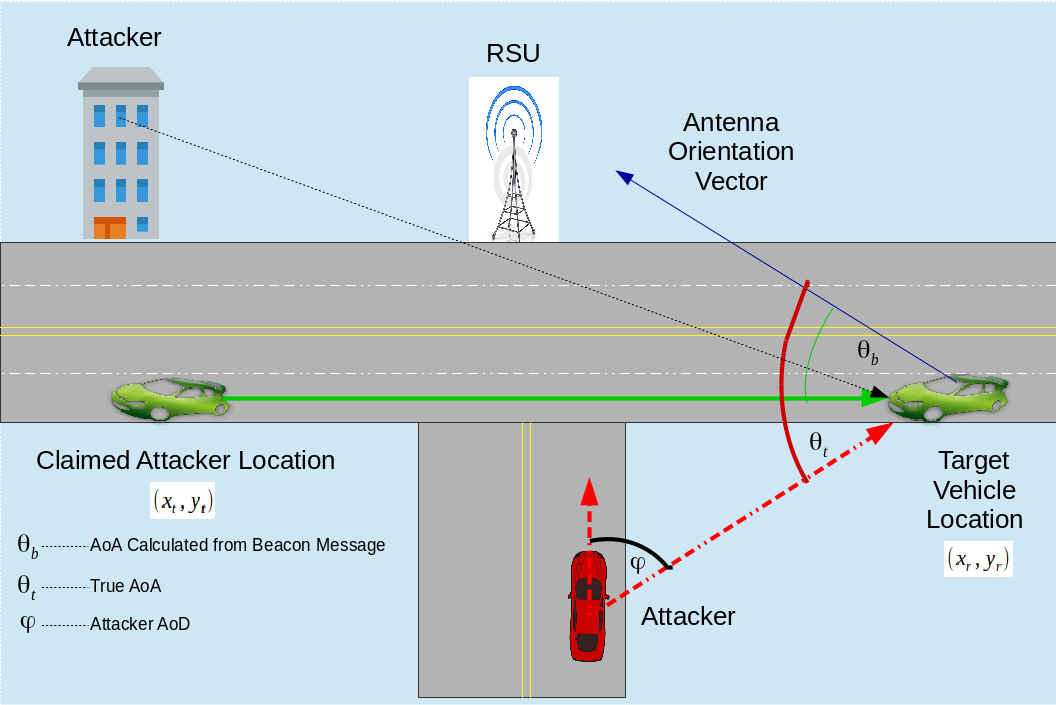}
\caption{System Model
\label{fig:sysmodel}}
\end{figure}

\begin{align}
\label{eq:inout} 
 {\mathbf{Y}[i,j,k]} &= {\mathbf{H}}[i,j,k] \mathbf{X}[i,j,k] + \mathbf{N}[i,j,k],
\end{align}
where $1 \leq i \leq n_p , 1 \leq j \leq n_s$ and $1 \leq k \leq 64$ denotes the packet number, symbol index and subcarrier index respectively, $\mathbf{X}[i,j,k] \in \mathbb{C}^{n\times 1}$ is the attacker signal constrained by an instantaneous maximum power constraint $\mathbb{E}\left[\mathbf{tr}\left(\mathbf{X}[i,j,k]\mathbf{X}^{\dagger}[i,j,k]\right)\right] \leq \mathbf{P}$. Also, ${\mathbf{H}}[i,j,k] \in {\mathbb{C}^{n \times n}}$ is the channel coefficients matrix between attacker and target vehicle. Finally, $\mathbf{N}[i,j,k] \in \mathbb{C}^{n\times1}$ is an independent zero mean circular symmetric complex random vector, $\mathbf{N}\sim \mathcal{CN}(0,R_{N})$ where $R_{N} = \sigma_N^2 \mathbf{I}_{n}$.
\par
VANETs are designed to provide wireless access for vehicles in a line of sight (LOS) environment for a maximum distance of 1 Km. Thus, we model the channel as a Rician fading channel. In Rician fading model, the received signal can be decomposed into two components; one is the specular component originated from the LOS path and the other is the diffuse component due to ground reflections and scatters from neighboring vehicles and other objects in the environment, or generally the non-line of sight component (NLOS). The LOS component can be considered fixed while the NLOS component can be best described as a Rayleigh fading channel. 
\begin{align}
\label{eq:ric_decomp}
\mathbf{H} = \mathbf{H}^{\footnotesize{\text{LOS}}} + \mathbf{H}^{\footnotesize{\text{NLOS}}}, 
\end{align}
where $\mathbf{H}^{\footnotesize{\text{LOS}}}$ and $\mathbf{H}^{\footnotesize{\text{NLOS}}}$ represents the LOS and NLOS components respectively and
\begin{align}
\label{eq:ric_decomp_det}
\mathbf{H}^{\footnotesize{\text{LOS}}} &= \sqrt{\dfrac{k}{1+k}} \left(\dfrac{1}{\sqrt{2}}+\dfrac{j}{\sqrt{2}}\right) {\mathbf{\Psi}} \nonumber \\
\mathbf{H}^{\footnotesize{\text{NLOS}}}  &= \sqrt{\dfrac{1}{2(1+k)}}\hat{\mathbf{H}}, 
\end{align}
where $k$ is the Ricean factor, ${\mathbf{\Psi}} =\mathbf{a}_r(\theta)\mathbf{a}_t^{\dagger}(\phi)$, $\mathbf{a}_r(\theta)$ and $\mathbf{a}_t(\phi)$ are the antenna array steering vectors at receiver and transmitter respectively, $\theta$ and $\phi$ are the AoA and AoD of the transmitted signal respectively as shown in Fig. (\ref{fig:sysmodel}). $\hat{\mathbf{H}} \sim \mathcal{CN}(\mathbf{0},\mathbf{I}_{n \times n})$ represents the channel coefficients matrix for the NLOS signal component.

 For the {ULA} configuration, the entries of the steering vectors are given by 
\begin{align}
\label{eq:steering}
  &\mathbf{a}(\theta) =
 \begin{bmatrix}
     1 &  z & z^2 & \dots & z^{n-1}
\end{bmatrix}^T \nonumber \\
&z = e^{-j2\pi\dfrac{d\sin(\theta)}{\lambda}},
 \end{align}
where $\lambda$, $d$, and $n$ are the wavelength of the center frequency of the transmitted signal, array elements spacing and size respectively. We parametrize the contribution of the NLOS and LOS components to the signal with $\sigma = \sqrt{{1}/{2(1+k)}}$, $\mu = \sqrt{k/(1+k)}$, respectively and choose $\mu^2 + 2\sigma^2 = 1$ for simplicity. It worth mentioning that, AWGN and Rayleigh fading channels are in fact limiting cases of the Rician fading channel.

\subsection{Conventional PKI Message Authentication in VANET}
\label{sec:conv_pki}
\begin{figure}[!htb]
\centering
\includegraphics[width=2.6 in, height = 1.9 in]{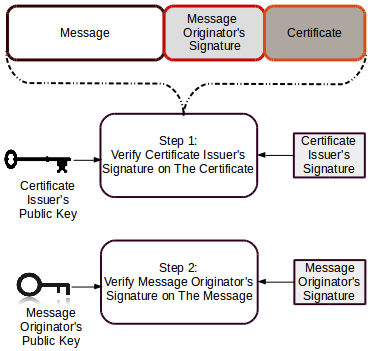}
\caption{Typical PKI Message Authentication in VANET
\label{fig:pki_auth}}
\end{figure}

In this section we give basics of PKI based message authentication in VANETs. As illustrated in Fig. (\ref{fig:pki_auth}), the message originator sends its certificate with its signed message to the receiver. The message receiver first verifies the certificate issuer's signature on the certificate using the public key of the CA. A successful verification indicates that the public key on the certificate belongs to the subject of the certificate. The message receiver proceeds to use this public key to verify the signature on the received message. A successful verification informs the message receiver that the message was signed by the subject of the certificate and that the message content has not been altered since it was signed. Each sender must be sure that all the receivers have its certificate before they need to verify its signature.

 \subsection{Attack Model}
\label{subsec:attack_model}
\textbf{Attacks on message authentication}. As noted before, we study message authentication problem in VANETs under the following attack scenarios;
\begin{enumerate}
\item Attacker with single or multiple stolen identities (Sybil attack) that intentionally falsies different GPS locations for different identities.
\item Attacker that spoofs/jams GPS information \cite{gps_spoof,gps_spoof_2} of other nodes in the network to let them report their incidents in certain/erroneous GPS locations. 
\end{enumerate}
The attacker associate the stolen certificate, $\mathtt{C}$, to its payload message, $\mathtt{M}$, signed with its private key, $\mathbf{priv}_{a}$. Thus the concatenated attacker message can be expressed as
\begin{align}
\mathfrak{M} = <ID\;|\;\mathtt{M}\;|\;\mathbf{sig}\left[\mathtt{M},\mathbf{priv}_{a}\right]\;|\;T|\;\mathtt{C}>,
\end{align}  
where $ID$ is the stolen identity, $\mathbf{sig}\left[\mathtt{M},\mathbf{priv}_{a}\right]$ is the signature of the attacker using its private key, $T$ is the time stamp with $'|'$ as the message concatenation operator. Further, the attacker is assumed to have the freedom to declare false GPS location information to achieve its goal. 
\par
Noting that in the considered attack scenario, message received at the target receiver will pass the conventional PKI procedures as it is either originated from an attacker with a stolen identity or a legitimate transmitter with spoofed/jammed GPS information. To the best of the authors knowledge, no solutions are available at the upper security layer for attacks with stolen identities or attacks with spoofed or jammed GPS information.

\textbf{Attack on SK exchange}. Let two legitimate vehicles, namely $\mathbf{A}$ and $\mathbf{B}$, attempt to share a secret key, $K$, using PK based SK agreement protocol without a trusted third party. The attacker, $\mathbf{E}$, (again with a stolen identity) is considered to be an active eavesdropper which is able to intercept the communication between $\textbf{A}$ and $\mathbf{A}$ and attempts to establish independent connections with both legitimate vehicles simultaneously and relay messages between them in order to let them believe they are communicating directly to each other over a secure link. In such scenario, the attacker has to perform two independent successful impersonation attacks with both legitimate vehicles to successfully taking over the control of the overall communication link. Doing so, the attacker will be able to intercept, decode, modify or even fabricate all relayed messages. This attack is known as the \textit{Man-in-the-Middle} (MitM) attack \footnote{In practice, the MitM attack is more applicable to wired rather than wireless networks due to the untethered nature of the wireless medium. However, a successful MitM attack may be accomplished by reactively sending a jamming signal once link activity is detected. Hence, only the attacker will be able to hear the transmitting node message and form an appropriate response while the victim receiver is kept ignorant by the effect of the attacker jamming signal.  Moreover, the attacker may try to perform an impersonation attack with either or each node proactively rather than reactively in order to either inject a false information or to reveal sensitive information a victim node may have.} where it combines \textit{Impersonation}, \textit{Modification/substitution} and/or \textit{Fabrication} attacks simultaneously.

\section{Basic Limits of AoA Estimation}
\label{sec:limits}
In section \ref{sec:joint}, we will introduce the AoA information as a security parameter that is used in conjunction with conventional PKI as solution to the problem of security against attackers with stolen identities. Therefor, we start by introducing the fundamental limits of AoA estimation. We first note that there is no prior distribution assumed for $\theta$, therefore, the associated estimation problem is non-Bayesian. In estimation theory, the Cramer Rao Bound (CRB) sets an upper bound on any parameter estimation performance. In particular, it defines the lower bound of the best estimator variance in terms of the solution of the following problem:
\begin{align}
  \label{eq:rx_obj_nojam}
  &\min_{\substack{\hat{\theta}(\cdot) \\\mathbb{E}\left[\mathbf{tr}\left(\mathbf{X}[i,j,k]\mathbf{X}^{\dagger}[i,j,k]\right)\right] \leq \mathbf{P}}}  \;\; \textbf{var}\left(\hat{\theta}\left(\mathbf{Y}\right)\right). 
 \end{align} 
To evaluate the CRB, we start by introducing
\begin{align}
\label{eq:eq_noise}
\mathbf{Z}[l] = \mathbf{H}^{\footnotesize{\text{NLOS}}}[l]\mathbf{X}[l] + \mathbf{N}[l],
\end{align}
where $\mathbf{Z}$ incorporates all undesired interfering components of the received signal, $1 \leq l \leq L$ where $L=4\times n_s \times n_p$ is the total number of pilot subcarriers contained in a message consists of $n_p$ packets each packet contains $n_s$ symbols. Since the receiver objective is to estimate the AoA of the LOS component, the NLOS diffuse component originated from ground reflections or scatters from neighboring vehicle is also considered as an undesired signal. Note that $\mathbf{Z}[l] \sim \mathcal{CN}\left(\mathbf{0}_{n \times 1},\mathbf{R}_z[l]\right)$. 
 Accordingly, the posterior distribution of the observation $\mathbf{Y}$ is given as follows:
 \begin{multline}
 \label{eq:posterior_with_jam}
f_{\mathbf{Y}|{H}_t^{\footnotesize{\text{LOS}}},\mathbf{X},} (\mathbf{y}) = \dfrac{1}{\prod_{l=1}^{L} \det(\pi\mathbf{R}_{z}[l])} \\
\times \exp \bigl\{-\dfrac{1}{L}\sum_{l=1}^{L}(\mathbf{y}[l]-{H}^{\footnotesize{\text{LOS}}}[l]{\mathbf{X}}[l])^{\dagger}\mathbf{R}_{z}[l]^{-1}  \\
(\mathbf{y}[l]-{H}^{\footnotesize{\text{LOS}}}[l]{\mathbf{X}}[l])\bigr\},
\end{multline}
which yields the following log-likelihood function
\begin{align}
\mathcal{L}(\mathbf{y}) &= - \sum_{i=1}^{L} \ln \det(\pi\mathbf{R}_{z}[l]) \nonumber \\  
& - \mathbf{tr}\left(\mathbf{R}_{z}^{-1}(\mathbf{y}-{H}^{\footnotesize{\text{LOS}}}{\mathbf{X}})(\mathbf{y}-{H}^{\footnotesize{\text{LOS}}}{\mathbf{X}})^{\dagger}\right), 
\end{align}
where $\mathbf{R}_{z} = \dfrac{1}{L}\sum_{i=1}^{L}\mathbf{R}_{z}[l]$. Further, It can be shown that, the CRB of AoA estimation is given by
\begin{align}
 \label{eq:crb_stoica}
 \mathbf{CRB} &= \dfrac{1}{2}\left[\sum_{l=1}^{L}\mathbf{Re}\left(\mu\mathbf{X}^{\dagger}[l] \hat{\mathbf{D}}^{\dagger} \mathbf{G}(\theta)\hat{\mathbf{D}}\mathbf{X}[l]\mu\right)\right]^{-1}  \nonumber \\
 &= \dfrac{1+k}{2Lk \mathbf{P} \hat{\mathbf{D}}^{\dagger} \mathbf{G}(\theta)\hat{\mathbf{D}}}, 
\end{align}
where
\begin{align}
&\hat{\mathbf{D}} = \mathbf{R}_{z}^{-1/2}\mathbf{D} \nonumber \\
&\mathbf{D}=\partial{\mathbf{a}}/\partial \theta \nonumber \\
&\mathbf{G}(\theta)  = [\mathbf{I} - {\mathbf{a}}({\mathbf{a}}^{\dagger}{\mathbf{a}})^{-1}{\mathbf{a}}^{\dagger}] \nonumber
\end{align}
where the dependence of $\mathbf{a}$ on $\theta$ was dropped for ease of notation. We note that, as $k \rightarrow \infty$, only LOS component is present and $\mathbf{R}_{z} \rightarrow \sigma_N^2\mathbf{I}_{n}$. Also, one can show that (this was also discussed in \cite{AOA_KNOWN}) consistent estimator exists in the large sample limit ($L \rightarrow \infty$) whereas efficient estimator exists only asymptotically in the array size ($n \rightarrow \infty$) \cite{AOA_SEC}. Moreover, the ML-AoA estimator given by:
 \begin{align}
 \label{eq:ml_aoa}
\hat{\theta}(\mathbf{Y}) &= \argmin_{\theta} \mathbf{tr}\left(\mathbf{B}^{\dagger}\mathbf{R}_{z}^{-1/2} \mathbf{G}(\theta)  \mathbf{R}_{z}^{-1/2} \mathbf{B}\right)
\end{align} 
achieves the CRB with equality asymptotically in the large array size limit and is a consistent estimator, where
\begin{align}
\label{eq:defs}
&\mathbf{B} = \mathbf{R}_{xy}^{\dagger} \mathbf{R}_{xx}^{-1} \;\;\; \in \mathbb{C}^{n \times n}, \\
&\mathbf{R}_{xy} = \dfrac{1}{L}\sum_{l=1}^{L}\mathbf{X}[l]\mathbf{Y}^{\dagger}[l] \;\;\;  \in \mathbb{C}^{n \times n},  \nonumber \\
&\mathbf{R}_{xx} = \dfrac{1}{L}\sum_{l=1}^{L}\mathbf{X}[l]\mathbf{X}[l]^{*} \;\;\;  \in \mathbb{C}^{n \times n}. \nonumber
\end{align} 
Furthermore, note that the regularity conditions required for the normality of the ML-AoA estimator \cite{asymptotic,asymptotic_normal} holds for the likelihood function (\ref{eq:posterior_with_jam}) of the considered model. Thus, in the limit of large sample, the ML-AoA estimator converges in distribution to a random variable with a truncated \footnote{The truncation in the normal distribution is due to the finite support of the ML-AoA estimator. We limit the support $\theta$ to the interval $[-\pi/2,\pi/2]$ due to the ULA antenna configuration, however, for 2-D antenna configuration with $360^\circ$ resolution, the support of $\theta$ is extended to $[-\pi,\pi]$.} normal distribution by the central limit theorem with mean equals to the true AoA and variance equals to the CRB given in Eq. (\ref{eq:crb_stoica}). This can be formally expressed as follows:
\begin{align}
\label{eq:ml_dist}
f_{\hat{\theta}}(\hat{\theta}(\mathbf{Y})) = \dfrac{1}{\sqrt{2\pi \mathbf{CRB}}} \dfrac{\exp\{\dfrac{-(\hat{\theta}-\theta)^2}{2\mathbf{CRB}}\}}{Q\left(\dfrac{-\pi/2 - \theta}{\sqrt{\mathbf{CRB}}}\right) - Q\left(\dfrac{\pi/2 - \theta}{\sqrt{\mathbf{CRB}}}\right)},
\end{align}
where $Q$ is the tail probability of the standard normal distribution. The results obtained in this section will be useful in the subsequent analysis in the rest of this paper.

\section{Joint Physical-security layer Message Authentication}
\label{sec:joint}
In this section, we delve into the details of the proposed physical layer assisted message authentication scheme. In the scope of VANETs, the exchanged safety and warning messages, which usually comes in the form of incident reporting, is highly dependent on the location of the message originators. Thus, vehicles in VANETs are required to periodically exchange beacon messages that include speed, time and location information. These information can be used to predict the physical direction of a given message source based on its GPS location claimed in its beacon message, $\theta_b$. In this work we propose that, an estimate for the AoA, $\hat{\theta}$, can be formed to be cross validated with one computed at the upper layers. The major advantage of this method is to provide the upper communication layers with a potential awareness about the physical communication environment. Having such advantage, security protocols at the upper communication layers are enabled with physical information that can help in a more informed security decision.
\par 
In what follows, by expected AoA we mean the AoA calculated from the GPS location information claimed in the beacon message of a given transmitter and denote it $\theta_b$. Meanwhile, by the estimated AoA, denoted $\hat{\theta}$, we mean the AoA estimate obtained from the physical layer signal representation of a given transmitter.

\subsection{Proposed Physical layer Assisted Authentication} 
\label{sec:proposed} 
\begin{figure}[!tb]
\centering
\includegraphics[width=3.2 in, height = 2.7 in]{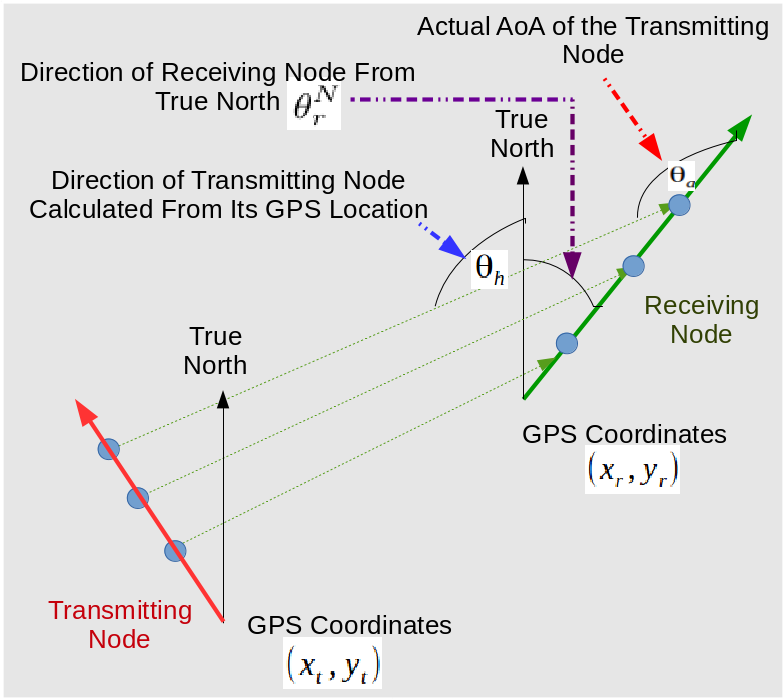}
\caption{The relation Between the Estimated AoA and the bearing information calculated from the GPS location Information.
\label{fig:AOA_Estimation_Diff_Orien}}
\end{figure}

We assume all vehicles/RSUs are associated with anonymous identity $ID$, private key, $\mathbf{priv}$, public key, $\mathbf{pub}$, and granted a certificate $C$ encrypted using the CA private key. Each vehicle/RSU is assumed to have GPS location information about itself and about other neighboring vehicles/RSUs using the periodically exchanged beacon messages. In the first place, the received message will go through the conventional PKI message authentication procedures as described in Section \ref{sec:conv_pki}. In the security layer, GPS location information can be used to calculate the angle of the transmitting node measured from true north. Thus, the receiving node should expect the transmission of a transmitting node at a given GPS location on an AoA equals to the calculated angle measured from true north taking into consideration the direction of its own antenna array orientation vector. The direction of the antenna array orientation vector is defined by the direction of the antenna array axis to the true north. Fig. (\ref{fig:AOA_Estimation_Diff_Orien}) illustrates the relation between the expected AoA and the bearing information that can be calculated from the exchanged GPS location information.
\par
Let the pairs $(x_t,y_t)$ and $(x_r,y_r)$ be the longitude and latitude coordinates of the transmitting and receiving nodes respectively. Then, the heading angle, $\theta_h$, measured from the true north of a plane wave emitted at $(x_t,y_t)$ and received at $(x_r,y_r)$ can be calculated as follows:
\begin{align}
\label{eq:head_ang}
\theta_h = atan2(\nu,\upsilon),
\end{align}
where $atan2$ is the arctangent function with two arguments \cite{atan2} and
\begin{align}
\label{eq:defs_heading}
\nu &= \cos(y_r) \sin(x_r - x_t), \nonumber \\
\upsilon &= \cos(y_t) \sin(y_r) - \sin(y_t) \cos(y_r) \cos(x_r - x_t).
\end{align}
\par
Denoting the angle between the receiver antenna array axis to the true north by $\theta_r^N$, then, the receiver can calculate the expected AoA, $\theta_b$, of that particular transmitter according to the information in its beacon message as follows:
\begin{align}
\label{eq:exp_aoa}
\theta_b = \theta_h + \theta_r^N.
\end{align} 
Further, at the physical layer, an estimate $\hat{\theta}$ is formed according to Eq. (\ref{eq:ml_aoa}) for the actual AoA of the received message. Using the estimate $\hat{\theta}$ from the physical layer and the expected AoA arrival, $\theta_b$, each vehicle/RSU builds up a table containing the expected as well as the estimated AoAs of the transmission of each of the other neighboring vehicles/RSUs as illustrated in Table (\ref{Table:aoa_list}).
\begin{table}[!h]
\centering
 \begin{tabular}{||c |c| c | c ||} 
 \hline
 Vehicle ID   & GPS Location & Expected AoA & Estimated AoA\\ [0.5ex] 
 \hline\hline
$ID_1$ & $(x_1,y_1)$ & $\theta_b^1$ & $\hat{\theta_1}$\\
\hline
$ID_2$  & $(x_2,y_2)$ & $\theta_b^2$ & $\hat{\theta_2}$\\
\hline
\vdots   & \vdots & \vdots & \vdots\\
\hline
$ID_i$ & $(x_i,y_i)$ & $\theta_b^i$ & $\hat{\theta_i}$\\
\hline
\vdots   & \vdots & \vdots & \vdots\\
\hline
$ID_{n_v}$ & $(x_{n_v},y_{n_v})$ & $\theta_b^{n_v}$ & $\hat{\theta_{n_v}}$\\ [0.5ex]
 \hline
\end{tabular}
\caption{List of Expected and Estimated AoAs Associated to Each Vehicle/RSU ID}
\label{Table:aoa_list}
\end{table}
Note that, those messages that fail in PKI message authentication procedures will be dropped and it will not have an entry in the above table. Thus, $n_v$ is the number of vehicles/RSUs in the vicinity that passed the PKI authentication procedures successfully. Further, each vehicle keeps updating the values of Table (\ref{Table:aoa_list}) whenever a message is received. Now, the receiver objective is to check the consistency between the declared AoA of a given message source and the estimate of the actual AoA. In the next section, we develop mathematical formulation of the receiver objective as a hypothesis testing problem.

\subsection{Formulation of AoA Authorization as a Hypothesis Testing Problem}
\label{sec:hyp_test}
Based on the knowledge of the declared AoA of arrival of all neighboring vehicles/RSUs extracted from the beacon messages, $\theta_b$, the problem of deciding whether the received signal is originated from the declared physical direction of a given vehicle/RSU can be formulated as a two sided hypotheses testing problem as follows:
\begin{align}
\label{eq:hyp_test}
\mathcal{H}_0 &: \theta \in \Gamma_0  \nonumber \\
\mathcal{H}_1 &: \theta \in \Gamma_1,
\end{align} 
where $\Gamma_0$ and $\Gamma_1$ are the decision regions for $\mathcal{H}_0$ and $\mathcal{H}_1$ receptively and are defined as follows:
\begin{align}
\label{eq:hyp_test_2}
&\Gamma_0 =  [-\pi/2,\theta_b) \cup (\theta_b,\pi/2] \nonumber \\
&\Gamma_1 =  \{\theta_b\}.
\end{align}
Note that, the probability of misdetection, $P_{MD}$, which is the probability of rejecting a true $\mathcal{H}_1$ hypothesis, corresponds to denying a transmission originated from the legitimate transmitter. Whereas, the false alarm probability of accepting a false $\mathcal{H}_0$ hypothesis will correspond to \textit{impersonation} probability as access will be granted to an illegitimate transmitter.
\par
Recalling the posterior distribution of the received signal given in Eq. (\ref{eq:posterior_with_jam}), we observe that $\mathbf{Y}|\mathcal{H}_i \sim \mathcal{CN}(H_{\theta}^{LOS}\mathbf{X},\mathbf{R}_u)$ where $\theta \in \Gamma_i$ and $i \in \{0,1\}$. Among different hypothesis testing techniques like Likelihood Ratio (LR), Lagrange Multiplier (LM) or Score tests, the Wald test is the most convenient test for the considered hypothesis testing problem. That is due to the highly nonlinear relation between the observation, $\mathbf{Y}$, and the composite parameter we test for \cite{wald}, $\theta$ in our case. The Wald test statistics can be found as
\begin{align}
\label{eq:wald}
\dfrac{|\hat{\theta} - \theta_b|}{\sqrt{\mathbf{CRB}}} \quad\mathop{\gtreqless}_{\mathcal{H}_1}^{\mathcal{H}_0}\quad \alpha,
\end{align}
where, $\hat{\theta}$ is the ML-AoA estimator given in Eq. (\ref{eq:ml_aoa}) and $\alpha$ is the decision threshold. Note that, the CRB of AoA estimation is a function of the AoA as can be seen in Eq. (\ref{eq:crb_stoica}) with the fact that angles near the array axis, $-\pi/2$ or $\pi/2$, experience much higher CRB than those close to zero. Thus we can notice that, the Wald test statistics accounts for that problem by incorporating the CRB to achieve an adaptive decision threshold. Recalling the distribution of the ML-AoA estimator given in Eq. (\ref{eq:ml_dist}), we define both the probability of detection an probability of false alarm as follows:
\begin{align}
\label{eq:pf_pf}
P_D = P\left(\dfrac{|\hat{\theta} - \theta_b|}{\sqrt{\mathbf{CRB}}} \leq \alpha | \mathcal{H}_1\right) \nonumber \\
P_F = P\left(\dfrac{|\hat{\theta} - \theta_b|}{\sqrt{\mathbf{CRB}}} \leq \alpha | \mathcal{H}_0\right)
\end{align}

 Thus, one would expect a relatively high false alarm probability as the attacking node approaches in a close vicinity to the declared AoA, $\theta_b$. 

\section{Secret Key Agreement Protocol}
\label{sec:SK_AGREE}
\begin{figure}[!htb]
\centering
\includegraphics[width=2.7 in, height = 2.6 in]{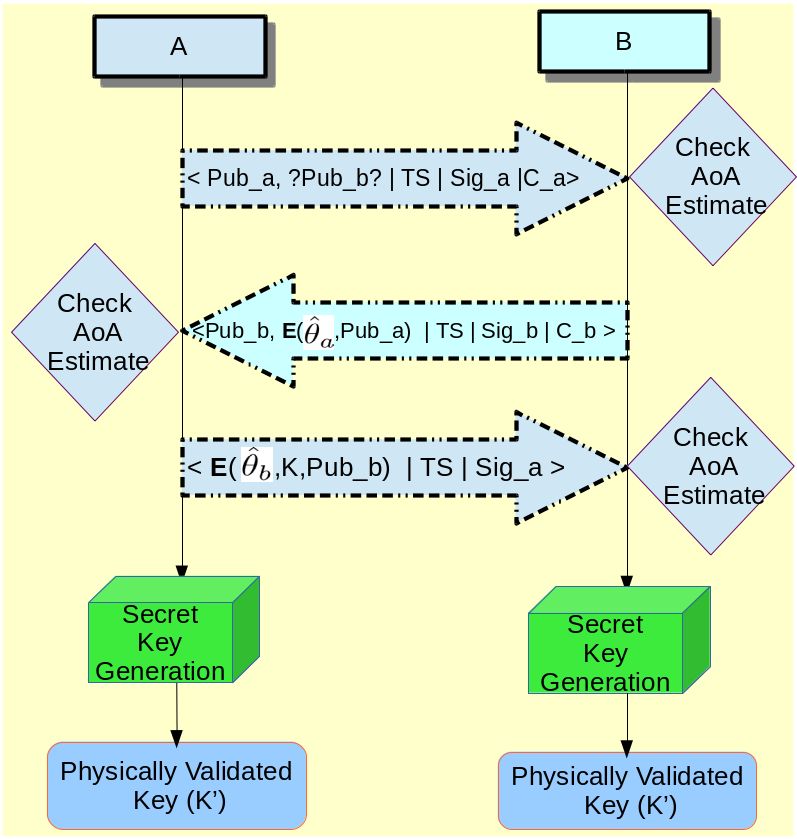}
\caption{Procedures of the Proposed Physical layer Assisted Secret Key Agreement Protocol.
\label{fig:SK_AG_PROCEDURES}}
\end{figure}

\begin{figure}[!htb]
\centering
\includegraphics[width=2 in, height = 2 in]{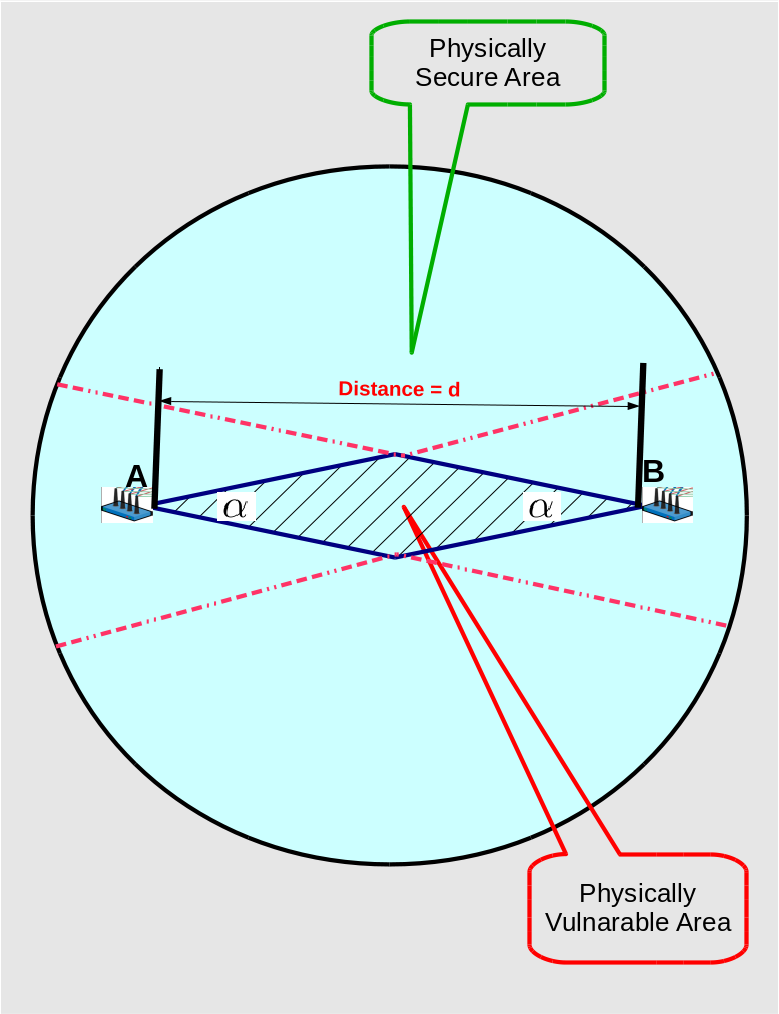}
\caption{Dividing the communication environment into physically vulnerable area and secure area.
\label{fig:sectors}}
\end{figure}

VANET is a vehicular wireless enabling technology that is expected to enable peer vehicles to establish a secure communication link without interaction with a trusted third party. In contrast to public key cryptosystems, symmetric key cryptosystems offers the advantage of low communication overhead as well as relatively low computational complexity. However, symmetric key cryptosystems require transmitting and receiving communication vehicles to agree on a secret key prior to communication. The process of sharing the secret key is called Secret Key Agreement. In this section, we provide a novel physical layer assisted secret key agreement protocol in which communicating vehicles are able to validate their respective physical location based on the claimed location information in the beacon messages. We consider a public key based secret key agreement between vehicles $\mathbf{A}$ and $\mathbf{B}$ without trusted third party interaction. As shown in Fig. (\ref{fig:SK_AG_PROCEDURES}), the proposed SKA procedures can be illustrated in the following steps
\begin{enumerate}
\item $\mathbf{A}$ select one of its preloaded public/private key pairs together with their corresponding certificate. 
\item $\mathbf{A}$ sends a message including its own public key asking for $\mathbf{B}$'s public key.
\item $\mathbf{B}$ accepts the message if it both pass the conventional PKI authentication and the AoA estimate, $\hat{\theta}_a$, is consistent with claimed location information, otherwise, abort.
\item $\mathbf{B}$ sends a message including its own public key together with an $m$ bits quantized version of the estimate $\hat{\theta}_a$ encrypted with $\mathbf{A}$'s public key.
\item $\mathbf{A}$ accepts the message if it both pass the conventional PKI authentication and the AoA estimate, $\hat{\theta}_b$, is consistent with claimed location information, otherwise, abort.
\item $\mathbf{B}$ sends a message including secret session key, $K$, together with an $m$ bits quantized version of the estimate $\hat{\theta}_b$ encrypted with $\mathbf{B}$'s public key.
\item $\mathbf{B}$ accepts the message if it both pass the conventional PKI authentication and the AoA estimate, $\hat{\theta}_a$, is consistent with claimed location information, otherwise, abort.
\item Both $\mathbf{A}$ and $\mathbf{B}$ employs an arbitrary function $\mathfrak{g}(K,\hat{\theta}_a,\hat{\theta}_b)$ to generate the physically validated session key $K'$.
\end{enumerate}
Note that in all of the propose SKA procedure, physical location validation is crucial for moving into the next step. This physical layer awareness introduces the concept of physical hardness to an attacker performing MitM attack. Now, it is clear that the literal meaning of the word "middle" became mandatory as such attacker has to be be in the middle between target vehicles. Otherwise, its AoA will not be validated and the impersonation attempt will fail. This significantly limits the region of possible locations for the potential MitM attackers to initiate their attack. In Figure (\ref{fig:sectors}), we introduce the concept of area security where a successful MitM attack requires the attacker to be physically located in the physically vulnerable area shown in figure.


\par
The proposed physical layer assisted secret key agreement protocol provides the following advantages
\begin{enumerate}
  \item It provides physical awareness of the environment over which the authenticated key agreement takes place.
  \item Secret key exchange protocols without a trusted third party are susceptible to  MitM attack, however, with the proposed algorithm, a successful Man in the  Middle attack would require the attacker to be physically located on the LOS path between the communicating nodes.
\end{enumerate}

\section{Simulation Results}
\label{sec:sim}
\begin{figure}[!htb]
\centering
\includegraphics[width= 3.5 in, height = 1.4 in]{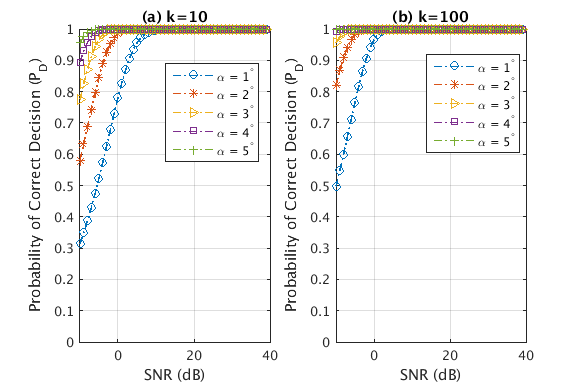}
\caption{$P_D$ as a function of $SNR$ for $\alpha=1^{\circ} : 5^{\circ}$ for a vehicle located at $25^{\circ}$. (a) $k=10$. (b) $k=100$.
\label{fig:PD}}
\end{figure}

\begin{figure}[!htb]
\centering
\includegraphics[width= 3.5 in, height = 1.4 in]{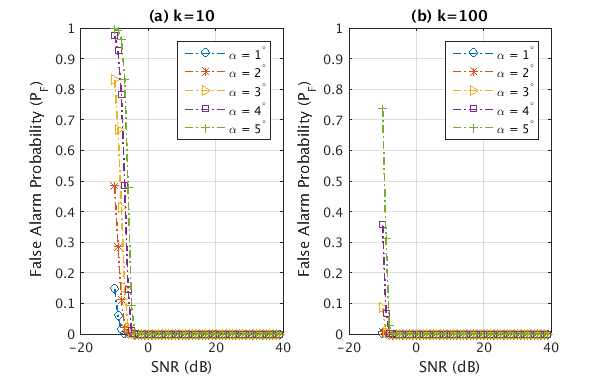}
\caption{$P_F$ as a function of $SNR$ for $\alpha=1^{\circ}:5^{\circ}$ for an attacker located at $-35^{\circ}$ with a claimed location of $-25^{\circ}$. (a) $k=10$. (b) $k=100$.
\label{fig:PF_FAR}}
\end{figure}

The simulation results provided in this section is based on the following simulation setup:
\begin{description}
\item[$\bullet$] \textbf{802.11P Protocol Stack.} A MATLAB implementation for the 802.11p standard was implemented as defined in \cite{802_11_STD} except that we have added MIMO capability.
\item[$\bullet$]\textbf{Array Size.} All communication nodes are of array size $n=4$.
\item[$\bullet$]\textbf{Training vector.} The predefined training sequence is acquired from the pilot tones embedded in the protocol stack.
\item[$\bullet$]\textbf{Payload Messages.} Payload messages are generated from a zero mean, unit variance complex Gaussian random variable and scaled to satisfy the power constraint.
\item[$\bullet$]\textbf{Communication Channels.} All communication channels are generated according to Equations (\ref{eq:ric_decomp}) and (\ref{eq:ric_decomp_det}). The entries of the channel matrix of the Rayleigh part of the channel are generated from a zero mean, unit variance complex Gaussian random variable and then scaled each by the corresponding value of $\sigma$. Meanwhile, the LOS component is generated for different values of $k$ which we will mention for each simulation scenario.
\end{description}
\textbf{Physical layer Assisted Message Authentication.} We evaluate the proposed physical layer assisted message authentication in terms of the probability of detection, $P_D$, and the false alarm probability, $P_F$ as a function of signal to noise ratio $SNR$ for decision threshold $\alpha$ ranges from $1^{\circ}$ to $5^{\circ}$ an Ricean factor $k=10$ and $k=100$. While both $P_D$ and $P_F$ are functions of the legitimate and attacker locations respectively, we shall give three different scenarios:
\begin{enumerate}
\item To evaluate $P_D$, legitimate vehicle located at $25^{\circ}$ is considered. As shown in Fig. (\ref{fig:PD}.(a)), $P_D$ approaches $1$ as $SNR$ and $\alpha$ increase, meanwhile, the same result holds true for increasing the Ricean factor from $k=10$ to $k=100$ of course with $P_D$ approaches $1$ faster as shown in Fig. (\ref{fig:PD}.(b)). 
\item To evaluate $P_F$, we first consider attacker vehicle located  at $-35^{\circ}$ while it claims that it is located away from its claimed location, namely at $-25^{\circ}$. As shown in Fig. (\ref{fig:PF_FAR}.(a)), $P_F$ approaches $0$ faster for smaller values of $\alpha$ as $SNR$ increase, meanwhile, the same result holds true for increasing the Ricean factor from $k=10$ to $k=100$ of course with $P_F$ approaches $0$ faster as shown in Fig. (\ref{fig:PF_FAR}.(b)).
\item Then, we consider attacker vehicle located  at $40^{\circ}$ while it claims that it is located at $37.5^{\circ}$. 
\end{enumerate}
   
 \begin{figure}[!htb]
\centering
\includegraphics[width= 3.5 in, height = 1.4 in]{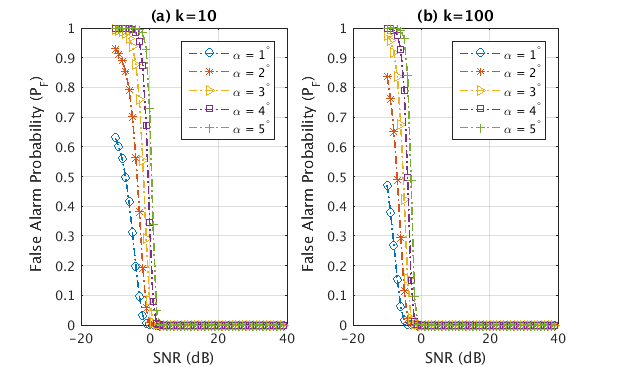}
\caption{$P_F$ as a function of $SNR$ for $\alpha=1^{\circ}:5^{\circ}$ for an attacker located at $40^{\circ}$ with a claimed location of $37.5^{\circ}$. (a) $k=10$. (b) $k=100$.
\label{fig:PF_CLOSE}}
\end{figure}
  
\begin{figure}[!htb]
\centering
\includegraphics[width= 3.5 in, height = 1.8 in]{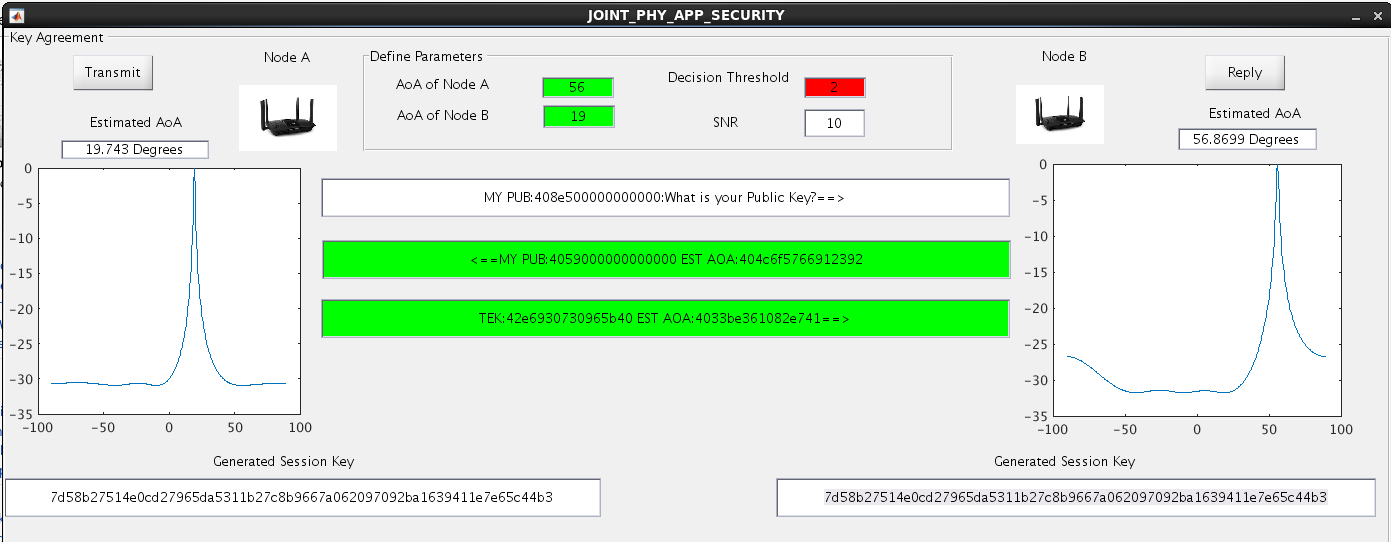}
\caption{Matlab GUI Implementation for the Proposed SKA, AoA is checked for correctness at each message, $\alpha=2$, $SNR=10dB$.
\label{fig:SKA_GUI}}
\end{figure}

\textbf{Physical layer Assisted SKA.} Fig. (\ref{fig:SKA_GUI}) provides a graphical user interface (GUI) developed in MATLAB in order to visualize the SKA protocol procedures given in section \ref{sec:SK_AGREE}.

\section{Discussion and Conclusion}
Communication security in Vehicular Ad-hoc Communication Networks (VANET) is in a direct relation to public safety. This work leverages the possible potential cooperation between the traditional security protocols used in VANETs and information about the physical environment that can be estimated at the physical layer. Taking into account the inherited location dependence of safety messages exchanged over VANETs, direction of arrival is a physical property that can be exploited to amplify trust level of a given message source. 
\par
In this paper, a message source authentication scheme for VANETs is proposed that is based on the cooperation between the traditional PKI authentication procedures in VANETs together with the signal direction of arrival estimated at the physical layer. Based on the periodically exchanged beacon messages, our scheme compares the AoA estimate of the signal observed at the physical layer with the "claimed" AoA based on the location information contained in the safety message. It is shown that, the provided security gain comes with no extra communication overhead, bandwidth or transmit power as opposed to security solutions provided in the upper layers.
\par
We also provided a novel physical layer assisted secret key agreement protocol in which communicating vehicles are able to validate their respective physical location based on the claimed location information in the beacon messages. We showed that the risk of the MitM attack, which is common in PKI SKA protocols with no trusted third party, is waived up to the literal meaning of the word "middle". 

\bibliographystyle{IEEEtran}
\bibliography{IEEEabrv,References}
%
%
%

\end{document}